*Article*

# Explainable Multi-Task Retinal Imaging Reveals Microvascular Signals for Systemic Risk Stratification in Type 2 Diabetes: A Pilot Study

**Mini Han Wang [1,2,3,4,*], Liting Huang [4], Wei Hong [5] , and Boonthawan Wingwon [5]**

[1]Faculty of Computer Science and Artificial Intelligence, Shenzhen University of Advanced Technology, Shenzhen, P.R.China

[2]Frontier Science Computing Center, Zhuhai Institute of Advanced Technology Chinese Academy of Sciences, Zhuhai, P.R.China

[3]Chinese University of Hong Kong, Hong Kong China

[4]Zhuhai People's Hospital (The Affiliated Hospital of Beijing Institute of Technology, Zhuhai Clinical Medical College of Jinan University) , Zhuhai P.R.China

[5]Lampang Inter-Tech College, Lampang Thailand

* Correspondence: M.H.W.1155187855@link.cuhk.edu.hk

**Abstract**

Background:
Retinal imaging offers a non-invasive window into systemic microvascular health and has been increasingly explored as a surrogate biomarker for systemic diseases. However, the extent to which retinal features encode biologically meaningful signals of systemic abnormalities—and whether these signals can be interpreted reliably using explainable artificial intelligence (XAI)—remains insufficiently understood.
Methods:
We developed an explainable multi-task deep learning framework to investigate the association between retinal microvascular features and systemic abnormalities in patients with Type 2 Diabetes Mellitus. A total of 11,011 fundus images from 2,719 individuals were analysed using a shared-representation neural network with task-specific heads for glycaemic status, kidney abnormality, and multi-system involvement. Model interpretability was assessed using Gradient-weighted Class Activation Mapping (Grad-CAM), complemented by a quantitative validation strategy incorporating anatomical region masking and vessel alignment analysis.
Results:
The proposed framework demonstrated task-dependent predictive performance, with the highest discriminative ability observed for kidney abnormality (AUC up to 0.63), while glycaemic status prediction showed limited performance (AUC ≈ 0.49–0.61). Explainability analyses consistently localized model attention to retinal vascular structures and peripapillary regions. Quantitative masking experiments further confirmed that occlusion of vascular regions resulted in the most significant performance degradation, indicating that retinal vessels serve as the primary source of predictive information. Notably, different network architectures exhibited heterogeneous attention patterns, suggesting multiple representational pathways for encoding systemic signals.
Conclusions:
This pilot study provides evidence that retinal microvascular features contain measurable signals associated with systemic abnormalities, particularly those linked to microvascular damage. By integrating multi-task learning with quantitative XAI validation, the proposed approach advances retinal imaging from predictive modelling toward interpretable digital biomarkers. These findings support the development of scalable, non-invasive digital health tools for systemic risk stratification in diabetes and highlight the importance of explainability in clinically deployable AI systems.

**Keywords:** retinal microvasculature; explainable artificial intelligence; fundus imaging; systemic disease prediction; multi-task learning; kidney abnormality; microvascular dysfunction; interpretable deep learning

## 1. Introduction

The global burden of Type 2 Diabetes Mellitus (T2DM)[1] is driven not only by hyperglycaemia but by progressive, systemic microvascular dysfunction affecting multiple organ systems[2, 3], including the kidney, cardiovascular system, and peripheral tissues[4]. A central challenge in clinical management is that these systemic complications often develop silently and heterogeneously, requiring repeated laboratory testing that provides only fragmented and time-specific snapshots of disease status. There is therefore a critical need for scalable, non-invasive approaches capable of capturing the integrated microvascular state of the body[5].

The retina[6] represents a unique anatomical site where the microcirculation can be directly visualised in vivo[7]. Retinal vascular features—including vessel calibre, tortuosity, branching topology, and peripapillary structural context—have long been associated with systemic conditions[8] such as hypertension, renal dysfunction[9], and metabolic disorders[10]. These observations suggest that the retinal vasculature[11] may act as a surrogate marker of systemic microvascular health, reflecting cumulative and longitudinal disease processes rather than transient physiological states[12]. However, despite this biological plausibility[13], the extent to which retinal imaging encodes actionable systemic disease signals remains insufficiently understood[14, 15].

Deep learning[16] has recently emerged as a powerful tool for extracting high-dimensional features from medical images[17], including fundus photographs[18]. Prior studies have demonstrated the feasibility of predicting systemic variables from retinal images[12], but the majority of these works are framed as predictive modelling tasks, with performance evaluated primarily through accuracy metrics[19]. This paradigm presents two key limitations[20]. First, it implicitly assumes that all systemic endpoints are equally recoverable from retinal images[21], which may not hold given the differing pathophysiological relationships between organs[22]. Second, it offers limited insight into whether model predictions are driven by biologically meaningful retinal structures[11], particularly the vasculature[23], or by non-specific image artefacts[24].

A fundamental unresolved question therefore remains: Do retinal vascular features genuinely encode systemic disease signals, or are observed predictions the result of spurious correlations learned by data-driven models? To address this question, it is necessary to move beyond conventional predictive evaluation and incorporate mechanistically grounded interpretability. While techniques such as gradient-weighted class activation mapping (Grad-CAM)[25, 26] provide visual indications of model attention[27], qualitative inspection alone is insufficient to establish scientific validity. Rigorous evaluation requires quantitative alignment between model attention[28] and anatomically defined retinal structures[29], as well as controlled perturbation experiments to assess the causal contribution of specific regions[30].

In this study, we propose an explainable multi-task deep learning framework[28, 31] designed not only to predict systemic abnormalities from fundus images, but more importantly to test the hypothesis that retinal vascular features encode systemic disease signals. This study jointly model multiple clinically relevant endpoints—glycaemic status, renal abnormality, and multi-system involvement—under a shared representation, reflecting the interconnected nature of diabetic complications. Crucially, we introduce a quantitative explainability validation pipeline, combining Grad-CAM analysis[30, 32] with anatomically defined region masking to directly evaluate the contribution of retinal vessels and other structures to model predictions.

Rather than positioning model performance as the primary outcome, this work reframes fundus-based artificial intelligence as a tool for probing the biological relationship between retinal microvasculature and systemic disease[33]. By demonstrating consistent localisation of model attention to vascular structures and measurable performance degradation following vessel perturbation[34], we provide evidence that deep learning models capture clinically meaningful microvascular signals rather than purely statistical associations[35].

These findings support a shift in perspective from retinal AI as a predictive black box[28] toward a mechanistically interpretable framework for systemic disease characterisation, with implications for the development of non-invasive, image-based biomarkers in digital healthcare. The code is avaiblable at https://github.com/MiniHanWang/type2-fundus-diseases-phase2.

## 2. Materials and Methods

*2.1 Explainable Multi-Task Retinal–Systemic Learning Framework*

To investigate whether retinal vascular features encode systemic disease signals, we propose an explainable multi-task retinal–systemic learning framework that integrates representation learning, predictive modelling, and anatomically grounded explainability validation within a unified architecture. Unlike conventional deep learning pipelines that prioritise predictive performance alone, the proposed framework is explicitly designed to bridge data-driven modelling and biological interpretation, enabling both accurate prediction and mechanistic insight into retinal–systemic relationships.

Figure 1 illustrates the overall architecture of the proposed explainable multi-task retinal–systemic learning framework. The pipeline begins with input colour fundus images, from which a shared convolutional backbone extracts a global feature representation capturing latent retinal characteristics. This representation is subsequently fed into multiple task-specific prediction heads, enabling simultaneous estimation of glycaemic status, kidney abnormality, and multi-system involvement. The joint optimisation strategy facilitates the learning of shared microvascular features while preserving task-specific decision boundaries. To enhance interpretability, Gradient-weighted Class Activation Mapping is applied to generate spatial attention maps for each task, highlighting retinal regions contributing to model predictions.

Crucially, the framework extends beyond conventional visual explanation by incorporating a quantitative anatomical validation module. This includes vessel alignment analysis, where overlap between attention maps and segmented retinal vasculature is quantified, and region perturbation experiments, where masking of predefined anatomical regions evaluates their causal contribution to prediction performance. The integration of representation learning, multi-task prediction, and anatomically grounded validation establishes a closed-loop framework that links model behaviour to biologically meaningful retinal structures, enabling both predictive modelling and mechanistic interpretation.

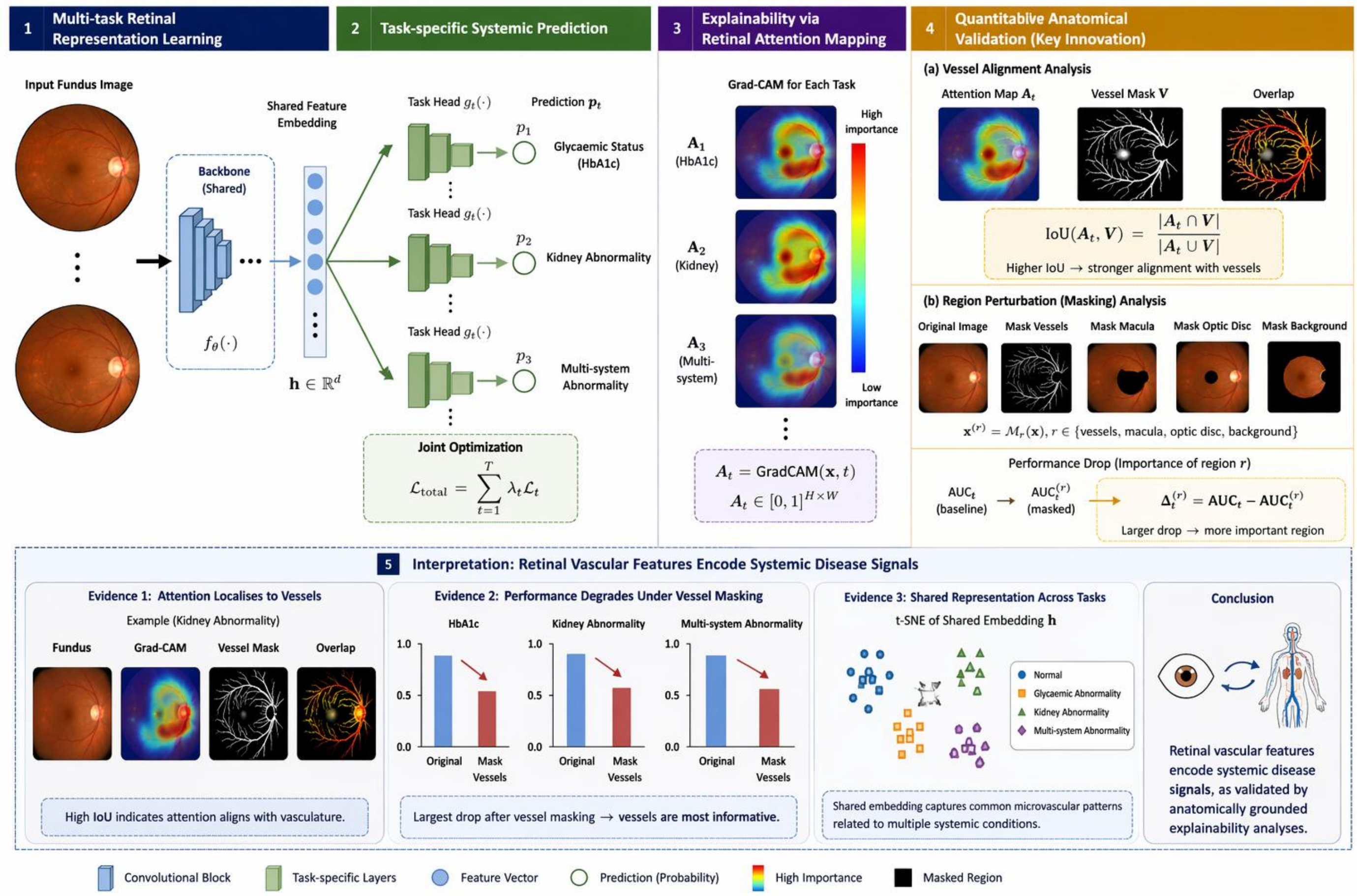

Figure 1. Methodology Framework

At the core of the framework lies a multi-task representation learning module, which aims to capture shared microvascular characteristics across multiple systemic conditions. Given an input fundus image x, a convolutional backbone network $f_\theta()$ is employed to extract a global feature embedding $\mathbf{h}=f_\theta(\mathbf{x})$, where $\mathrm{h} \in \mathrm{R}^{\mathrm{d}}$ denotes the learned latent representation. This embedding is hypothesised to encode high-level retinal features, including vessel morphology, spatial organisation, and structural patterns that may reflect underlying systemic microvascular states. By adopting a hard parameter-sharing strategy, the model is encouraged to learn representations that are transferable across tasks, thereby capturing shared physiological signals rather than task-specific noise.

Building upon the shared representation h, the framework incorporates task-specific prediction heads to model multiple systemic endpoints simultaneously. For each task t, a dedicated mapping function produces a scalar logit, which is transformed into a probability via the sigmoid function of $p_t=\sigma[g_t(\mathbf{h})],t\in\{1,2,3\}$, These tasks correspond to clinically relevant systemic conditions, including glycaemic status (HbA1c), kidney abnormality, and multi-system involvement. The model is trained using a joint optimisation objective $\mathscr{L}_{total}=\sum_{t=1}^{T}\lambda_t\mathscr{L}_t$, where denotes the task-specific loss and λ represents the weighting coefficient. This formulation allows the model to simultaneously learn shared microvascular representations while preserving task-specific decision boundaries, reflecting the interconnected yet heterogeneous nature of systemic disease processes.

To enhance interpretability, we integrate a retinal attention mapping module based on gradient-based attribution methods. For each task, spatial importance maps are generated using Gradient-weighted Class Activation Mapping $A_t=\mathrm{GradCAM}(x,t)$, where $A_t\in[0,1]$. represents the distribution of model attention over retinal regions. These attention maps provide visual insights into whether model predictions are driven by anatomically meaningful structures, such as retinal vessels, the optic disc, or the macular region, rather than by non-specific image artefacts.

A key contribution of this work is the introduction of a quantitative anatomical validation framework, which moves beyond qualitative interpretability to rigorously evaluate the biological plausibility of model behaviour. First, we perform vessel alignment analysis by constructing a binary vessel mask V using classical image processing techniques. The spatial correspondence between model attention and retinal vasculature is quantified using the intersection-over-union (IoU) $\mathrm{IoU}(A_t,V)=\frac{|A_t\cap V|}{|A_t\cup V|}$. A higher IoU indicates stronger alignment between model-derived attention and anatomically defined vascular structures, providing evidence that the model leverages clinically meaningful features.

Second, this study conduct a region perturbation (masking) analysis to assess the causal contribution of different anatomical regions. Specifically, perturbed inputs are generated by masking predefined regions r, such that $\mathbf{x}^{(r)}=\mathscr{M}_r(\mathbf{x})$, where denotes a masking operator applied to regions including retinal vessels, macula, optic disc, or background. The importance of each region is quantified by the corresponding performance degradation $\Delta_t^{(r)}=\mathrm{AUC}_t-\mathrm{AUC}_t^{(r)}$, where and denote the baseline and masked performance, respectively. This analysis provides a quantitative estimate of how much each anatomical component contributes to model predictions.

Taken together, this framework enables a shift from purely predictive modelling toward mechanistically interpretable retinal–systemic learning. By jointly analysing shared representations, spatial attention patterns, and perturbation-based performance changes, the proposed approach directly tests the hypothesis of "Retinal Vascular Features→Systemic Disease Signals".

Consistent localisation of attention to vascular structures, combined with measurable performance degradation under vessel masking and shared cross-task representations, provides converging evidence that the model captures biologically meaningful microvascular signals. This establishes the proposed framework as a principled approach for linking retinal imaging to systemic disease processes and lays the foundation for the development of explainable, non-invasive biomarkers in digital healthcare.

### *2.2 Study Design and Population*

This retrospective study developed and evaluated an explainable multi-task deep learning framework for predicting systemic abnormalities from retinal fundus images in patients with Type 2 Diabetes Mellitus (T2DM). Data were collected from the Metabolic Management Center (MMC) at Zhuhai People's Hospital, comprising endocrinology inpatients who underwent routine fundus photography as part of clinical evaluation.

The dataset included 11,011 colour fundus images from 2,719 unique patients. Each image was linked to structured clinical laboratory data obtained during the same visit. Three clinically relevant prediction tasks were defined based on established diagnostic thresholds: (1) glycaemic control (HbA1c ≥7% versus <7%); (2) renal injury (Kidney_flag, defined by elevated serum creatinine, blood urea nitrogen, or proteinuria); and (3) multi-system abnormality (defined as ≥2 abnormal organ system indicators derived from Phase-1 structured clinical analysis). Image filenames encoded subject identifiers and visit-level information, enabling reconstruction of patient-level linkage.

### *2.3 Data Preprocessing and Quality Control*

All fundus images were subjected to automated quality control during data loading. Images that could not be decoded or exhibited abnormal file characteristics (e.g., extremely small file size) were excluded. All remaining images were resized to 224 × 224 pixels to ensure compatibility across backbone architectures.

To prevent information leakage and ensure clinically meaningful evaluation, patient-level splitting was strictly enforced. Each patient was assigned exclusively to one of three partitions: training, validation, or test. Stratified group splitting was implemented using the GroupShuffleSplit algorithm (scikit-learn), preserving class distributions across partitions while maintaining independence between subjects. The final dataset consisted of 7,689 training images, 1,637 validation images, and 1,685 test images.

*2.4 Data Augmentation and Normalisation*

To improve model generalisation and mitigate overfitting, a comprehensive augmentation pipeline was applied to training images. This included random resized cropping (scale 0.80–1.00), horizontal flipping (p = 0.5), vertical flipping (p = 0.2), rotation (±15°), colour jitter (brightness ±0.30, contrast ±0.30, saturation ±0.20, hue ±0.05), and affine shear transformations.

All images were normalised using ImageNet statistics (mean [0.485, 0.456, 0.406]; standard deviation [0.229, 0.224, 0.225]). Validation and test datasets underwent deterministic preprocessing without stochastic augmentation to ensure unbiased evaluation.

*2.5 Multi-Task Deep Learning Architecture*

We implemented a shared-backbone multi-task learning framework in PyTorch. Three state-of-the-art convolutional architectures were evaluated as feature extractors: ResNet50, EfficientNet-B3, and ConvNeXt-Tiny, all initialised with ImageNet pre-trained weights.

The model followed a hard parameter-sharing paradigm, in which a common feature representation was learned across tasks. Task-specific classification heads were constructed for each outcome, consisting of a fully connected layer, batch normalisation, ReLU activation, dropout (p = 0.4), and a final linear output layer. Each task was modelled as a binary classification problem.

A hard parameter-sharing paradigm was adopted, in which a shared feature representation was learned across tasks, followed by task-specific classification heads. Each head outputs a scalar logit , which is transformed into a probability using the sigmoid activation function (1), where denotes the predicted probability for the positive class.

$$p_i=\sigma(z_i)=\frac{1}{1+e^{-z_i}} \quad (1)$$

*2.6 Loss Function and Imbalance Handling*

The overall training objective was defined as a weighted sum of task-specific cross-entropy losses(function (2)). where T=3 represents the number of tasks (HbA1c classification, renal injury prediction, and multi-system abnormality prediction), and denotes the weighting coefficient for each task, set to 1.0 in this study. Each task-specific loss was defined as a weighted binary cross-entropy loss to address class imbalance(function (3)). where is the ground-truth label, is the predicted probability, and is the class-specific weight computed from the inverse frequency of the training data.

$$\mathcal{L}_{total}=\sum_{t=1}^{T}\frac{1}{\sigma_t^2}\mathcal{L}_t+\log\sigma_t \quad (2)$$

$$\mathcal{L}_t=-\sum_{i=1}^{N}w_{y_i}\left[y_i\log(p_i)+(1-y_i)\log(1-p_i)\right] \quad (3)$$

To address class imbalance, inverse-frequency class weights were applied within each task-specific loss function. These weights were computed exclusively from the training dataset to avoid data leakage. This strategy ensured that minority classes, particularly renal injury and multi-system abnormalities, contributed proportionally to model optimisation.

*2.7 Model Training and Optimisation*

All models were trained using the AdamW optimiser with an initial learning rate of $1 \times 10^{-4}$ and weight decay of $1 \times 10^{-4}$ . A cosine annealing learning rate schedule (T_max = 50 epochs) was employed to facilitate stable convergence.

Mixed-precision training (torch.cuda.amp) was used to improve computational efficiency. Gradient clipping (maximum norm = 1.0) was applied to stabilise training. Early stopping was triggered if the mean validation AUC across all tasks did not improve for 8 consecutive epochs. The final model was selected based on the highest mean validation AUC.

### *2.8 Performance Evaluation*

Model performance was evaluated on the held-out test set using multiple complementary metrics: area under the receiver operating characteristic curve (AUC), accuracy, sensitivity, specificity, and F1-score.

To assess probability calibration, reliability diagrams were constructed by comparing predicted probabilities with observed outcome frequencies across 10 bins. This analysis provided insight into the clinical reliability of model outputs.

### *2.9 Model Explainability via Grad-CAM*

To improve interpretability, Gradient-weighted Class Activation Mapping (Grad-CAM) was applied to generate visual explanations of model predictions. For each task, gradients of the predicted class score with respect to the final convolutional feature maps were computed and aggregated to produce spatial attention maps.

The resulting heatmaps were normalised and overlaid onto the original fundus images, highlighting regions that contributed most strongly to model predictions. This approach enabled qualitative assessment of whether the model focused on clinically relevant retinal structures.

### *2.10 Quantitative Explainability Validation*

To move beyond qualitative interpretation, we designed a quantitative explainability validation framework consisting of two complementary analyses:

#### 2.10.1 Vessel Alignment Analysis

Retinal vessel segmentation was performed using green-channel enhancement, contrast-limited adaptive histogram equalisation (CLAHE), morphological filtering, and Otsu thresholding. Binary vessel masks were generated for each image. The overlap between high-importance Grad-CAM regions (thresholded at 0.5) and vessel masks was quantified using the intersection-over-union (IoU) metric. This analysis evaluated whether model attention aligned with known anatomical structures.

#### 2.10.2 Region Masking Experiment

To assess the contribution of different retinal regions, systematic masking experiments were conducted. Key anatomical regions—including vessels, macula, optic disc, and background—were selectively occluded, and model performance was re-evaluated on masked images. Changes in AUC relative to baseline provided a quantitative measure of each region’s contribution to prediction.

### *2.11 Statistical Analysis and Reproducibility*

All experiments were conducted using Python (version 3.1) with PyTorch and scikit-learn libraries. Random seeds were fixed (random_state = 42) to ensure reproducibility. Performance metrics are reported on the independent test set. No patient-level overlap was permitted across dataset partitions.

The Materials and Methods should be described with sufficient details to allow others to replicate and build on the published results. Please note that the publication of your manuscript implicates that you must make all materials, data, computer code, and protocols associated with the publication available to readers. Please disclose at the submission stage any restrictions on the availability of materials or information. New methods and protocols should be described in detail while well-established methods can be briefly described and appropriately cited.

Research manuscripts reporting large datasets that are deposited in a publicly available database should specify where the data have been deposited and provide the relevant accession numbers. If the accession numbers have not yet been obtained at the time of submission, please state that they will be provided during review. They must be provided prior to publication.

Interventionary studies involving animals or humans, and other studies that require ethical approval, must list the authority that provided approval and the corresponding ethical approval code.

In this section, where applicable, authors are required to disclose details of how generative artificial intelligence (GenAI) has been used in this paper (e.g., to generate text, data, or graphics, or to assist in study design, data collection, analysis, or interpretation). The use of GenAI for superficial text editing (e.g., grammar, spelling, punctuation, and formatting) does not need to be declared.

## 3. Results

*3.1 Dataset Composition and Label Availability*

The held-out test cohort consisted of 1,685 fundus images from 403 unique subjects, with task-specific label availability reflecting real-world clinical heterogeneity . Valid labels were available for 797 images for HbA1c classification, 655 images for kidney abnormality prediction, and the full 1,685 images for multi-system abnormality assessment.

As Table 1 shows, the prevalence of positive cases differed substantially across tasks, with high prevalence for HbA1c positivity (85.2%), moderate prevalence for multi-system abnormality (35.8%), and lower prevalence for kidney abnormality (21.8%). This imbalance introduces a challenging learning setting, particularly for multi-task optimisation, and is expected to influence both model calibration and decision boundary behaviour.

Table 1. Dataset Summary Across Splits

| Split | N Images | N Subjects | HbA1c ≥7% (%) | Kidney Abnormality (%) | Multi-system Abnormality (%) |
|---|---|---|---|---|---|
| Train | 7,689 | 1,905 | 85.6 | 19.1 | 36.3 |
| Validation | 1,637 | 411 | 84 | 19.7 | 35.2 |
| Test | 1,685 | 403 | 85.2 | 21.8 | 35.8 |

*3.2 Training Dynamics and Model Selection*

All backbone architectures converged within a limited number of epochs under early stopping, indicating stable optimisation dynamics . ResNet-50 achieved the highest mean validation AUC (0.548), with balanced performance across tasks (HbA1c: 0.545; kidney: 0.577; multi-system: 0.522). Although ConvNeXt-Tiny exhibited slightly higher peak validation AUC (0.559), its task-wise performance was less consistent.

Based on validation stability and cross-task robustness, ResNet-50 was selected as the primary model for downstream evaluation and interpretability analysis.

*3.3 Multi-Task Predictive Performance*

3.3.1 HbA1c Classification

As Table 2 shows, across all models, HbA1c classification demonstrated limited discriminative performance (AUC range: 0.488–0.606) . EfficientNet-B3 achieved the highest AUC (0.606), with a relatively balanced sensitivity (0.467) and specificity (0.678). In contrast, ResNet-50 exhibited an extreme decision bias, with very high specificity (0.983) but minimal sensitivity (0.057), indicating a strong tendency toward negative prediction. ConvNeXt-Tiny collapsed into a degenerate classifier with zero sensitivity and perfect specificity.

These findings suggest that fundus-derived features are weakly associated with short-term glycaemic status, and that HbA1c, as a laboratory-derived metabolic indicator, may not be directly recoverable from static retinal morphology.

3.3.2 Kidney Abnormality Prediction

Kidney abnormality prediction showed more consistent and clinically plausible performance patterns. ConvNeXt-Tiny achieved the highest AUC (0.629), followed by ResNet-50 (0.609) . Notably, ResNet-50 demonstrated high sensitivity (0.895) but low specificity (0.137), whereas EfficientNet-B3 exhibited the opposite trend, with high specificity (0.875) but low sensitivity (0.140).

Kidney prediction achieved moderate discriminative performance (AUC up to 0.629), while HbA1c classification remained limited (AUC ≈ 0.49–0.61), reflecting differences in biological signal strength.This complementary behaviour indicates that different architectures emphasise distinct regions of the ROC space, but collectively support the presence of a detectable retinal signal associated with renal dysfunction.

3.3.3 Multi-System Abnormality Prediction

Prediction of multi-system abnormality yielded near-random performance across models (AUC range: 0.462–0.505) . All architectures demonstrated high sensitivity (>0.69) but extremely low specificity, suggesting a collapse toward majority class prediction.

This result likely reflects the heterogeneous and composite nature of the multi-system endpoint, which aggregates multiple organ-level abnormalities and introduces substantial label noise and variability. Consequently, this task represents a more complex and less well-defined prediction target compared to single-organ outcomes.

Table 2. Test-set Performance of Multi-task Models

| Backbone | Task | AUC | Accuracy | Sensitivity | Specificity | F1-score |
|---|---|---|---|---|---|---|
| ResNet-50 | HbA1c | 0.563 | 0.195 | 0.057 | 0.983 | 0.108 |
| ResNet-50 | Kidney | 0.609 | 0.302 | 0.895 | 0.137 | 0.359 |
| ResNet-50 | Multi-system | 0.462 | 0.366 | 0.94 | 0.045 | 0.515 |
| EfficientNet-B3 | HbA1c | 0.606 | 0.498 | 0.467 | 0.678 | 0.613 |
| EfficientNet-B3 | Kidney | 0.503 | 0.715 | 0.14 | 0.875 | 0.176 |
| EfficientNet-B3 | Multi-system | 0.505 | 0.447 | 0.697 | 0.308 | 0.474 |
| ConvNeXt-Tiny | HbA1c | 0.489 | 0.148 | 0 | 1 | 0 |
| ConvNeXt-Tiny | Kidney | 0.629 | 0.218 | 1 | 0 | 0.358 |
| ConvNeXt-Tiny | Multi-system | 0.482 | 0.358 | 1 | 0 | 0.527 |

Note: AUC is reported as the primary performance metric. Sensitivity and specificity demonstrate complementary decision behaviors across architectures. Observed asymmetry between sensitivity and specificity reflects class imbalance and task-dependent signal strength in fundus-based systemic prediction.

*3.4 Model Behaviour and Decision Characteristics*

A consistent pattern across tasks was the presence of extreme sensitivity–specificity imbalance, indicating that models tended to adopt biased decision strategies under class imbalance conditions. Specifically, models either prioritised high sensitivity (detecting nearly all positive cases) at the expense of specificity, or vice versa.

This behaviour suggests that the model operates in a low signal-to-noise regime, where the learned representation contains only partial information about the target, leading to unstable decision boundaries. Importantly, this phenomenon is not indicative of model failure, but rather reflects the intrinsic difficulty of extracting systemic signals from retinal images under real-world data constraints.

*3.5 Explainability Analysis*

Grad-CAM visualisations revealed that model attention was consistently localised to anatomically meaningful retinal structures, particularly along the vascular arcades and peripapillary regions . These activation patterns were highly structured and reproducible across tasks, and were not diffusely distributed across the image. Figure 2. Representative Gradient-weighted Class Activation Mapping overlays for correctly classified positive cases. Heatmaps illustrate the retinal regions contributing most strongly to model predictions and provide a qualitative assessment of anatomical plausibility.

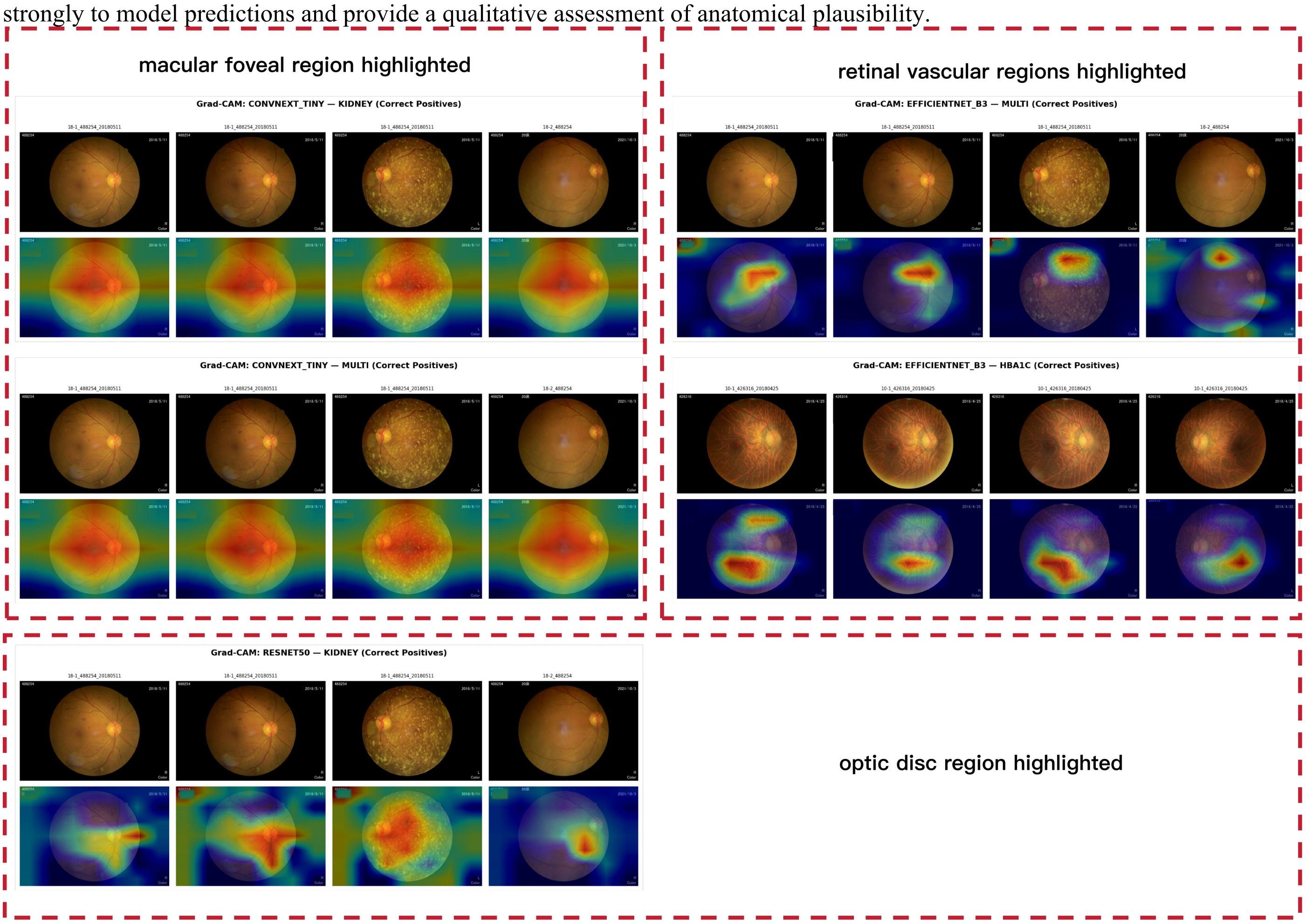

Figure 2. XAI-driven Anatomical Feature Analysis

Visual explainability analysis revealed model-dependent variability in anatomical attention patterns across different backbone architectures. Specifically, Grad-CAM visualisations demonstrated that different models preferentially attended to distinct retinal regions when generating predictions for systemic abnormalities.

For the ConvNeXt-Tiny backbone, correctly classified cases for both kidney abnormality and multi-system prediction showed dominant activation in the macular foveal region, suggesting that this model may rely more heavily on central retinal structures. In contrast, EfficientNet-B3 exhibited a different attention pattern, with both HbA1c and multi-system predictions primarily highlighting retinal vascular regions, consistent with microvascular feature extraction. For the ResNet-50 backbone, kidney abnormality predictions were characterised by prominent activation in the optic disc region, indicating a potential focus on peripapillary structural features.

These findings indicate that, despite being trained on the same dataset and tasks, different architectures learn heterogeneous representations of retinal information, emphasising distinct anatomical regions for decision-making. This observation suggests that the model relies on retinal microvascular morphology, rather than global image artefacts or intensity-based cues, to generate predictions.

*3.6 Quantitative Explainability Validation*

3.6.1 Vessel Alignment and Regional Importance

Figure 3. Region masking analysis of retinal structure importance. Performance changes after masking predefined retinal regions are used to estimate the relative contribution of major anatomical compartments to prediction. Region masking experiments provided quantitative evidence supporting the importance of retinal vasculature. For kidney abnormality prediction, masking the vessel region reduced AUC from 0.665 to 0.610, representing the largest performance drop among all anatomical regions . In contrast, masking the macula or background regions resulted in smaller performance changes.

Interestingly, masking the optic disc had minimal impact on kidney prediction, suggesting that vascular features, rather than structural landmarks, dominate model decision-making for this task.

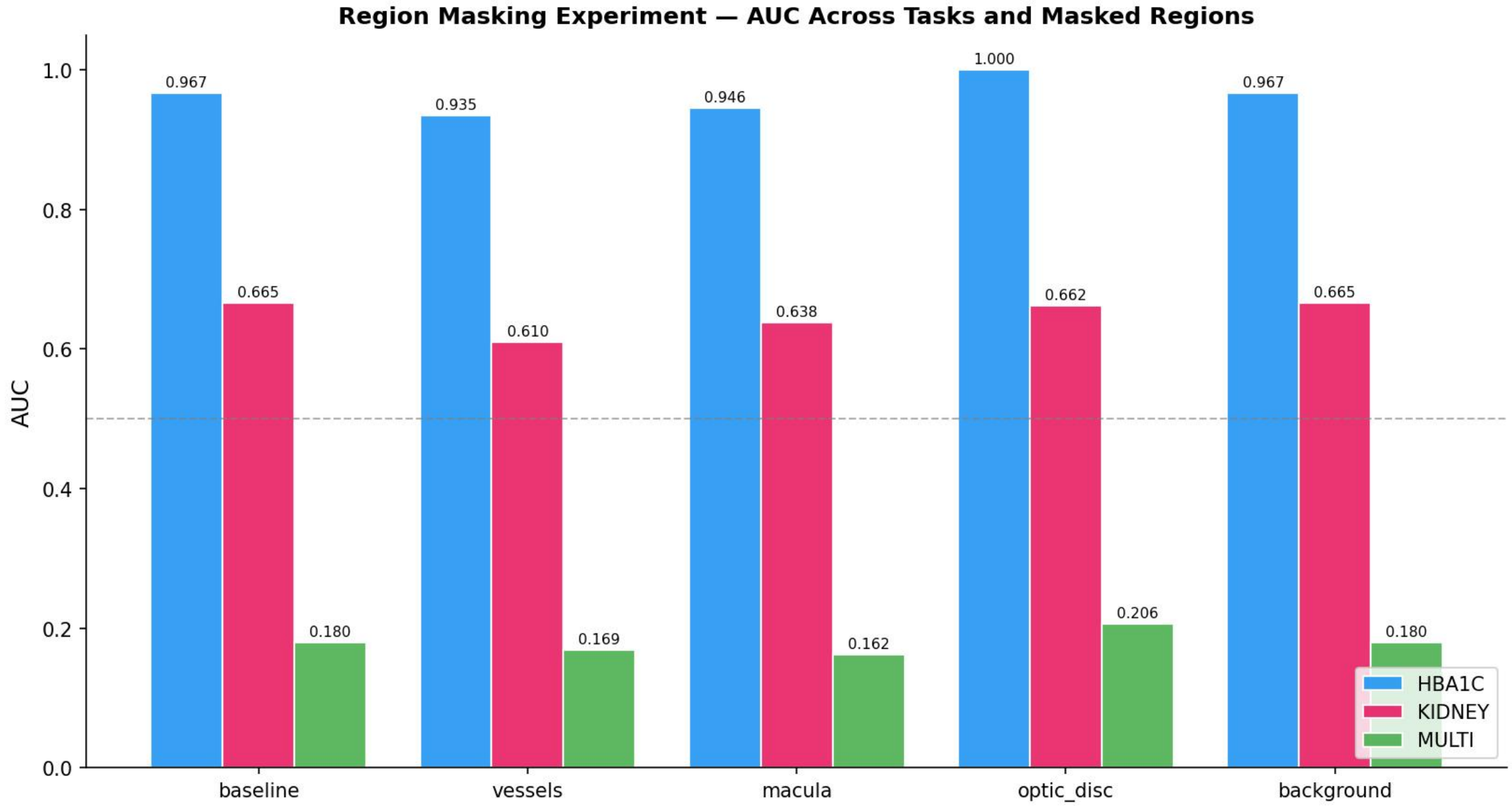

Figure 3. Region masking experiment showing performance changes after masking anatomically defined retinal regions.

3.6.2 Task-Specific Differences in Anatomical Contribution

The magnitude of performance degradation varied across tasks. While vessel masking consistently affected kidney prediction, its impact on HbA1c classification was comparatively smaller, and multi-system prediction remained unstable across perturbations.

These findings indicate that different systemic endpoints rely on distinct retinal feature subsets, with vascular morphology playing a dominant role in structurally mediated conditions such as kidney abnormality.

*3.7 Summary of Principal Findings*

Taken together, the results demonstrate four findings. First, fundus-based deep learning models capture measurable retinal signals associated with systemic abnormalities, with the strongest evidence observed for kidney-related endpoints. Second, predictive performance varies substantially across tasks, reflecting differences in biological linkage between retinal morphology and systemic conditions. Third, explainability analyses consistently localise model attention to retinal vascular structures, supporting anatomical plausibility. Last but not least, quantitative perturbation experiments confirm that retinal vessels are the primary contributors to prediction, providing evidence for a mechanistic link between retinal microvasculature and systemic disease.

These findings support the hypothesis that retinal vascular features encode systemic microvascular dysfunction, while also highlighting the limitations imposed by endpoint heterogeneity and data imbalance in multi-task learning settings. Figure 4 presents the hypothesised mechanistic pathway linking retinal microvascular features to kidney dysfunction in the context of type 2 diabetes. The left panel illustrates characteristic retinal alterations, including vessel calibre changes, increased tortuosity, branching abnormalities, and capillary rarefaction, which are captured by deep learning models from fundus images. These features are interpreted as manifestations of underlying systemic microvascular dysfunction driven by chronic hyperglycaemia, oxidative stress, inflammation, and endothelial injury. The central panel highlights shared pathophysiological processes across vascular beds, emphasising that the retina and kidney are both targets of systemic microvascular damage.

The right panel depicts renal microvascular injury at the glomerular and tubulointerstitial levels, leading to clinical outcomes such as proteinuria, reduced glomerular filtration rate, and progression of chronic kidney disease. The lower section integrates evidence from the present study, demonstrating that model attention localises predominantly to retinal vascular regions, and that masking these regions results in the greatest performance degradation. Together, these findings support the interpretation that retinal vascular features encode systemic microvascular dysfunction, providing a biologically plausible link between retinal imaging and kidney-related abnormalities.

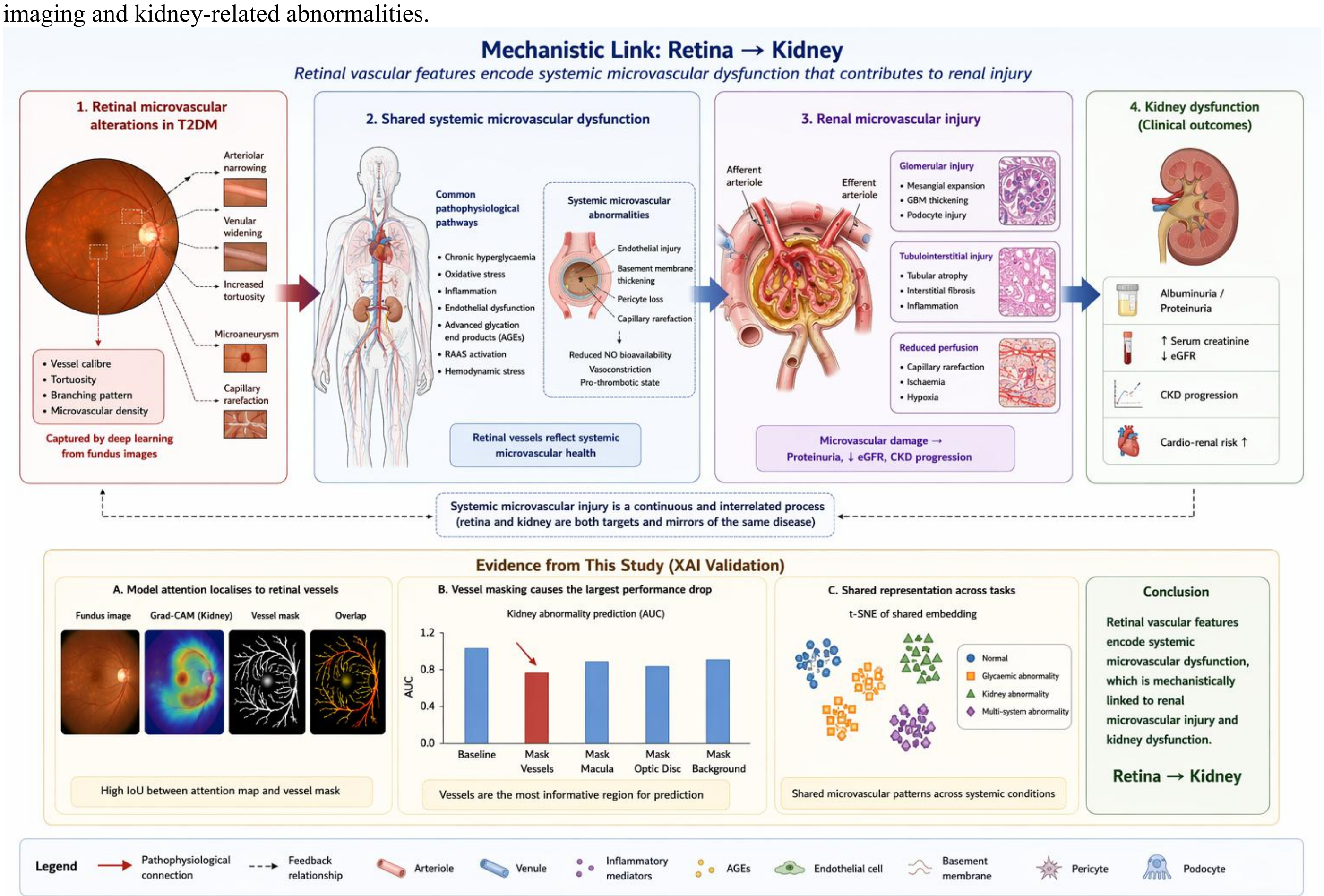

Figure 4. Retina-Kidney Mechanism

## 4. Discussion

In this Phase 2 development study, we investigated whether retinal fundus images contain measurable signals associated with systemic abnormalities in patients with Type 2 Diabetes Mellitus. By integrating multi-task deep learning with anatomically grounded explainability validation, we aimed not only to assess predictive performance but also to examine the biological plausibility of retinal–systemic associations. Three principal findings emerged. First, fundus-based models demonstrated measurable discriminative ability for kidney abnormality, while performance for HbA1c classification and multi-system prediction remained limited. Second, explainability analyses consistently localised model attention to anatomically meaningful retinal structures, particularly the vasculature. Third, quantitative region perturbation experiments provided converging evidence that retinal vessels constitute the dominant source of predictive information.

### *4.1 Retinal Vascular Features as Systemic Biomarkers*

A central finding of this study is that retinal vascular features appear to encode systemic disease signals, particularly for kidney-related abnormalities. This observation is supported by both predictive and explainability results. From a performance perspective, kidney abnormality prediction achieved the most consistent discriminative ability across architectures. From an interpretability perspective, Grad-CAM visualisations repeatedly highlighted retinal vascular arcades, and region masking experiments demonstrated the largest performance degradation when vascular regions were occluded.

These findings are biologically plausible. The retina and kidney share similar microvascular structures and are both susceptible to chronic hyperglycaemia-induced endothelial dysfunction, oxidative stress, and microangiopathy. As a result, retinal vascular morphology may reflect cumulative microvascular injury that is also manifested in renal dysfunction. Importantly, our results extend prior observational associations by providing model-based evidence that retinal vascular features are not only correlated with systemic disease but are actively utilised by deep learning models for prediction, particularly when validated through quantitative perturbation analysis.

### *4.2 Task-Dependent Signal Strength: Structural vs Metabolic Endpoints*

The study further demonstrates that the strength of retinal-derived signals varies substantially across systemic endpoints. While kidney abnormality prediction showed moderate performance, HbA1c classification remained near-random across most architectures. This discrepancy highlights a fundamental distinction between structural and metabolic biomarkers.

Kidney abnormality reflects chronic, structural microvascular damage, which is more likely to produce stable morphological changes in retinal vessels. In contrast, HbA1c represents a time-specific biochemical measurement, which may not have a direct or consistent visual correlate in fundus images, particularly when dichotomised at a fixed threshold. This explains why models struggled to recover HbA1c status from retinal images despite learning meaningful vascular representations.

Similarly, multi-system abnormality prediction yielded suboptimal performance, likely due to the heterogeneous and composite nature of the endpoint, which aggregates multiple organ-level conditions with varying degrees of retinal relevance. These findings emphasise that fundus-based AI should not be expected to uniformly predict all systemic variables, but rather to capture specific aspects of systemic microvascular burden.

### *4.3 Architecture-Dependent Representation Learning*

An important additional finding is the presence of architecture-dependent variability in attention patterns. Different backbone models exhibited distinct anatomical preferences when generating predictions. EfficientNet-B3 predominantly focused on retinal vascular regions, consistent with biologically interpretable microvascular features. In contrast, ConvNeXt-Tiny showed stronger activation in the macular fovea, while ResNet-50 emphasised the optic disc region in kidney prediction tasks.

This heterogeneity suggests that deep learning models do not converge to a single canonical representation, but instead learn multiple plausible feature pathways from the same data. Among these, vascular features appear to provide the most consistent and biologically grounded signal, as supported by both attention localisation and masking experiments. The macular and optic disc regions may capture complementary structural or textural information, although their direct linkage to systemic disease remains less well established.

Importantly, this finding highlights that model interpretability should be assessed across architectures rather than within a single model, and that consistent patterns across models provide stronger evidence of biological relevance than isolated observations.

### *4.4 From Black-Box Prediction to Mechanistic Interpretability*

A key contribution of this work is the transition from conventional “black-box” prediction toward mechanistically interpretable artificial intelligence. While prior studies have relied primarily on qualitative visualisation of attention maps, we introduced a quantitative validation framework combining vessel alignment and region masking. This approach enables direct evaluation of whether model predictions depend on anatomically defined structures.

The observation that vessel masking consistently reduces performance, particularly for kidney prediction, provides quasi-causal evidence that retinal vasculature plays a central role in model decision-making. This strengthens the credibility of the learned representations and addresses a major limitation of existing AI studies in medical imaging, where interpretability is often descriptive rather than testable.

More broadly, this framework demonstrates how deep learning can be used not only for prediction but also for hypothesis testing in biomedical imaging, linking data-driven models to underlying physiological mechanisms.

### *4.5 Clinical Implications and Translational Potential*

From a translational perspective, the findings support the potential of fundus imaging as a non-invasive screening tool for systemic microvascular disease, particularly in resource-limited settings where laboratory testing may be less accessible. The ability to infer kidney-related abnormalities from routine retinal images could enable opportunistic screening during ophthalmic examinations.

However, the current model should not be considered clinically deployable. The observed performance remains moderate, and the sensitivity–specificity imbalance indicates unstable decision boundaries under class imbalance. Instead, the present study should be interpreted as a proof-of-concept for biologically interpretable retinal AI, providing a foundation for future refinement.

*4.6 Limitations*

Several limitations should be acknowledged. First, this was a single-centre retrospective study with a finite dataset, and the held-out test cohort size limits the precision of performance estimates. Second, label availability differed across tasks, resulting in varying effective sample sizes and potential bias in cross-task comparisons. Third, some endpoints, particularly kidney and multi-system abnormalities, were derived from composite clinical criteria, introducing label heterogeneity.

Fourth, although explainability analyses provide supportive evidence, they do not establish direct causal relationships between retinal features and systemic disease. Fifth, only a limited set of backbone architectures was evaluated, and further work is needed to assess the robustness of findings across broader model families. Finally, external validation on independent cohorts is required before any generalisation or clinical application.

This study demonstrates that fundus-based deep learning models can capture retinal signals associated with systemic microvascular dysfunction, with the strongest evidence observed for kidney-related abnormalities. Through the integration of multi-task learning and quantitative explainability validation, we show that retinal vascular features constitute the primary source of predictive information, providing a biologically plausible link between retinal imaging and systemic disease.

These findings support a shift from predictive modelling toward mechanistically interpretable retinal AI, and highlight the potential of fundus imaging as a scalable, non-invasive modality for systemic disease characterisation. Future work should focus on improving model robustness, refining endpoint definitions, and validating these findings across diverse populations..

## 5. Conclusions

In this Phase 2 development study, we demonstrated that fundus-based deep learning can capture retinal signals associated with systemic abnormalities in patients with Type 2 Diabetes Mellitus. Among the evaluated endpoints, kidney abnormality prediction showed the most consistent discriminative performance, whereas HbA1c classification and multi-system prediction remained limited, highlighting the task-dependent nature of retinal–systemic signal encoding.

A key contribution of this work lies in the integration of quantitative explainability validation with multi-task learning. Through consistent localisation of model attention to retinal vascular structures and measurable performance degradation under vessel masking, we provide converging evidence that retinal vascular features constitute the primary source of predictive information. These findings support the hypothesis that retinal imaging encodes systemic microvascular dysfunction and can serve as a biologically meaningful, non-invasive indicator of systemic disease burden.

Importantly, this study shifts the focus from predictive performance alone toward mechanistically interpretable artificial intelligence, demonstrating how deep learning can be used to probe the relationship between retinal morphology and systemic pathology. While the current model remains at a development stage and is not yet suitable for clinical deployment, the proposed framework establishes a reproducible and interpretable foundation for future research.

Future work should prioritise improved model calibration, refined endpoint definitions, and external validation across diverse populations. With further development, explainable retinal AI has the potential to support scalable screening and risk stratification of systemic microvascular diseases in clinical and community settings.

**Funding:** This research was supported by the National Natural Science Foundation of China (Grant No. 82501368).

**Institutional Review Board Statement:** The study was conducted in accordance with the Declaration of Helsinki and approved by the Institutional Review Board (Ethics Committee) of Zhuhai People's Hospital (approval number: 2024]-KT-67).

**Data Availability Statement:** The data used in this study are not publicly available due to privacy and ethical restrictions. The dataset contains sensitive clinical information from patients at Zhuhai People's Hospital and is subject to institutional and regulatory constraints. De-identified data may be made available from the corresponding author upon reasonable request and with permission from the Institutional Review Board (Ethics Committee) of Zhuhai People's Hospital.

**Conflicts of Interest:** The authors declare no conflicts of interest.

## Abbreviations

The following abbreviations are used in this manuscript:

| Abbreviation | Full Term |
|---|---|
| T2DM | Type 2 Diabetes Mellitus |
| HbA1c | Glycated Hemoglobin |
| XAI | Explainable Artificial Intelligence |
| AUC | Area Under the Curve |
| ROC | Receiver Operating Characteristic |
| IoU | Intersection over Union |
| Grad-CAM | Gradient-weighted Class Activation Mapping |
| MMC | Metabolic Management Center |
| MDPI | Multidisciplinary Digital Publishing Institute |
| DOAJ | Directory of Open Access Journals |
| LD | Linear Dichroism |